\documentclass[preprint,showpacs,preprintnumbers,endfloats*,amsmath,amssymb]{revtex4-1}

\usepackage{graphicx}
\usepackage{dcolumn}
\usepackage{bm}
\newcommand{\vecvar}[1]{\mbox{\boldmath$#1$}}

\begin{document}

\preprint{PRESAT-8401}

\title{Real-space electronic-structure calculations with full-potential all-electron precision for transition-metals}

\author{Tomoya Ono, Marcus Heide}
\affiliation{Department of Precision Science and Technology, Osaka University, Suita, Osaka 565-0871, Japan}
\author{Nicolae Atodiresei, Paul Baumeister, Shigeru Tsukamoto, and Stefan Bl\"ugel}
\affiliation{Institut f\"ur Festk\"orperforschung and Institute for Advanced Simulation, Forschungszentrum J\"ulich and JARA, D-52425 J\"ulich, Germany}

\date{\today}

\begin{abstract}
We have developed an efficient computational scheme utilizing the real-space finite-difference formalism and the projector augmented-wave (PAW) method to perform precise first-principles electronic-structure simulations based on the density functional theory for systems containing transition metals with a modest computational effort. By combining the advantages of the time-saving double-grid technique and the Fourier filtering procedure for the projectors of pseudopotentials, we can overcome the egg box effect in the computations even for first-row elements and transition metals, which is a problem of the real-space finite-difference formalism. In order to demonstrate the potential power in terms of precision and applicability of the present scheme, we have carried out simulations to examine several bulk properties and structural energy differences between different bulk phases of transition metals, and have obtained excellent agreement with the results of other precise first-principles methods such as a plane wave based PAW method and an all-electron full-potential linearized augmented plane wave (FLAPW) method.
\end{abstract}

\maketitle
\section{Introduction}
Density functional theory\cite{dft} (DFT) in various approximations to the unknown exchange correlation potential developed to the most important practical scheme to investigate or determine the electronic, structural and many other properties of molecules, solids, and materials in physics, chemistry, materials science, mineralogy and reaches now into new fields such biophysics and electro-chemistry. A typical trend observed is that we deal with increasingly more complex systems characterized by many atoms of different chemical nature in open structures and of low symmetry. This path is supported by the increasing availability of increasingly powerful computers spearheaded by high-performance computers. The developments of the latter show that we have to cope with massively parallel computer architectures dealing with thousands of cores. This path accelerates in the next 10 years with the attempt moving from peta-scale to exa-scale computing. Thus, developing electronic structure methods whose applicability scales with the available processes becomes a prerequisite.

The real-space scheme of first-principles calculations, in which all computations are implemented in real space, is a method that has the potential to scale with massively parallel architectures and has this potential without compromise on the precision and thus should be superior to conventional plane wave methods\cite{beck}. Real-space methods can be loosely categorized as one of three types: finite differences\cite{chelikowsky1,chelikowsky2,icp}, finite elements\cite{white}, or wavelets\cite{wavelet}. Chelikowsky {\it et al.}\cite{chelikowsky1,chelikowsky2} have presented the real-space method combining the high-order finite-difference formula for the second derivative of the Kohn-Sham equation\cite{kohnsham} and the norm-conserving pseudopotential\cite{hsc,tm} and demonstrated its applicability for the investigation of the cohesive energy and bond length of diatomic molecules. There exist alternative high-order discretizations such as the Mehrstellen form used in the work of Briggs {\it et al.}\cite{briggs} Several techniques to improve the precision and accelerate the computational speed have been proposed so far\cite{tsdg1,tsdg2,icp,gygi,multigrid,quantchem}. Further advantages of real-space finite-difference (RSFD) formalisms are that (i) the computational costs involved in calculating the projectors of pseudopotentials can be reduced when the calculations are implemented in real space. (ii) Since all of the calculations are carried out in real space, it is easy to incorporate Wannier-type orbitals, which are localized in a finite region required for the realization of linear scaling calculations\cite{galli,dm}, into the algorithm. (iii) The grid spacing should be narrowed in order to improve the calculational precision, a procedure, which is simple and definite, and (iv) boundary conditions are not constrained to be periodic, e.g., combinations of periodic and nonperiodic boundary conditions for surfaces and wires, uneven boundary condition for triclinic systems, and twist boundary conditions for helical nanotubes are included straight forwardly\cite{ono-rsfd-appl}. In particular point (iv) is of significant advantage for electron-transport calculations\cite{icp,ono-transport}, because a nonperiodic boundary is indispensable for the direction in which electrons flow.

Norm-conserving pseudopotentials, which were introduced by Hamann {\it et al.}\cite{hsc}, have made significant contributions to the description of the band structure of semiconductors and simple metals and the computation of their bond lengths, crystal structures and surface reconstructions as well as vibrational modes of molecules\cite{ihm}. However, for systems with first-row elements or 3{\it d} electrons, the norm-conserving pseudopotentials are very hard so that a large plane-wave basis set or an extremely small grid spacing is required. Similarly, treating semicore states as valence states, which is often necessary for transition metals or compounds with light elements, e.g., GaN, results in hard pseudopotentials and affects their transferability. Some of these problems can be avoided employing Vanderbilt's ultrasoft pseudopotentials\cite{vanderbilt}, which relax the norm-conservation condition and are now adopted quite widely. Another alternatives is full-potential all-electron methods such as the Korringa-Kohn-Rostoker (KKR)-Greenfunction\cite{KKR} or the full-potential linearized augmented plane-wave (FLAPW) method\cite{flapw,fleur}. Both methods provide the Kohn-Sham answer of the given exchange correlation function taken for the problem at hand and provide in addition a precise treatment of wave functions near the nuclei probed by several experimental techniques and can supply properties that are usually not provided by the conventional pseudopotential approach. These are, among many others, the  hyperfine parameters and electric field gradients, but also correct magnetic structures. In particular, for simulations related to spintronics, the correct description of magnetism the precise treatment of the localized {\it d}-levels is of crucial importance. Bl\"ochl\cite{paw} proposed a state-of-the-art all-electron method, called the projector augmented-wave (PAW) method, that retains the formal simplicity and practicability of the traditional pseudopotential approach, but matches the precision of the full-potential all-electron methods. 

Mortensen {\it et al.}\cite{gpaw} implemented the PAW method into their RSFD computational code, Grid-based Projector Augmented-Wave (GPAW) code, and demonstrated that the code, in terms of computational efficiency, is comparable to a plane wave based PAW (PWPAW) method by computing the bond length and atomization energy of small molecules consisting of first- and second-row elements. In addition, they claimed that the average difference of atomization energies of small molecules from the results obtained by other computational codes is 50 meV. However, the RSFD calculations have never been applied in simulations that require extremely high precision such as comparisons of small structural energy differences between different bulk phases of transition metals. For example, the difference of the cohesive energy between face-centered-cubic (fcc) Cu and hexagonal close-packed (hcp) Cu is just 8 meV/atom according to the FLAPW calculation\cite{flapw,fleur} using the local density approximation\cite{VWN-lda} and the required precision for the discussion of a pressure-induced phase transition for CuPt from $L$1$_1$ to $B$2 is less than 10 meV\cite{cupt}.

In the RSFD formalisms, there is a well-known problem that the total energies and forces depend unphysically on the position of the nucleus relative to the positions of grid points, which is called the ``egg box effect.'' This problem is an obstacle to the precise computation of the total energies of systems containing transition metals with a moderate grid spacing. Although, reliable results can be achieved using a very small grid spacing in the RSFD formalisms, one of the greatest benefits of the PAW method, the precise treatment of transition metals with modest computational effort in CPU time and computer memory, will be lost and instead the computational cost is expected to increase substantially. Furthermore, in the case of PAW method, the grid spacing required for transition metals is smaller than that for first-row elements, while this relation is opposite in the case of the norm-conserving pseudopotentials. Thus, the egg box effect is more severe for transition metals in the case of the PAW method and it is of great importance to develop methods to circumvent the egg box effect. Several prescriptions to deal with it have been proposed over time and their applications proved successful within the framework of norm-conserving pseudopotentials up to now\cite{gygi,tsdg1,tsdg2,multigrid,quantchem}. However, as far as we know, there are no reports on efficient techniques that allow us to perform extremely precise simulations such as examinations of small structural energy differences of transition metals with a moderate grid spacing.

In this paper, we present an efficient computational scheme with a high degree of precision within the framework of RSFD formalisms making use of the PAW method to enable large-scale first-principles simulations for systems containing transition metals. By combining the advantage of the time-saving double-grid (DG) technique\cite{tsdg1,tsdg2} and Fourier filtering (FF) procedure for projectors of pseudopotentials\cite{king-smith}, we have succeeded to reduce the number of grid points employed in the calculations in the case of Cu 75\% compared to our previous procedure\cite{tsdg2} and the precision is improved. In order to demonstrate the performance of our scheme, we study the total-energy convergence with respect to the grid spacing and the energy variation due to the egg box effect. We also calculate the bulk properties of various 3$d$ transition metals and Cu, as well as the structural energy differences between different bulk phases of those. We compare these results to present that this RSFD formalism is very precise and that the precision is comparable to those to in-house calculations using the VASP code\cite{kresse-joubert}, which bases on the PWPAW method, and the FLEUR code\cite{fleur}, which uses the FLAPW method\cite{flapw}. The advantage of the in-house comparison is that we can use the same exchange correlation potential and the same pseudopotential and can converge the properties in question individually to achieve an accurate comparison. This comparison shows that indeed the RSFD formalism in combination with the PAW method is very precise and is able to achieve full-potential all-electron precision.

The rest of this paper is organized as follows: in Sec. II, we briefly introduce the trials for the egg box effect in the RSFD formalism and present our prescriptions to deal with this problem in detail. We introduce some examples to demonstrate the potential power of our scheme in Sec. III. Finally, in Sec. IV, we conclude with a discussion on the future direction of the RSFD electronic-structure calculations. 

\section{Methods}
\subsection{Egg box effect}
In the RSFD formalism, real-space grid points are distributed across the computational region in which the atoms are distributed and wave functions, electronic charge density, and potentials are all represented on the discrete grid points. The egg box effect describes the phenomenon that the total energies and forces are affected unphysically by the positions of the grid points relative to the nucleus, although their discretizations are not invariant under uniform translations of the system with respect to the position of the grid. This problem occurs also even in plane wave formalisms when the operations concerning the projectors of pseudopotentials are implemented in real space\cite{king-smith}. Although we can overcome the egg box effect by reducing the grid spacing, small grid spacings may require so many grid points as to result in a substantial increase in computational effort. It is known that the effect can be avoided by treating the projectors of pseudopotentials in Fourier space following the operation in conventional plane wave formalisms. However, this procedure results in an increase of the computational costs from $O(mM)$ to $O(mMN)$ and the degradation of the performance on massively parallel computers, where $m$, $M$, and $N$ are the numbers of atoms, occupied bands, and total grid points in the supercell, respectively. These procedures contradict an important demand in the simulations using PAW pseudopotentials, i.e., to perform very precise total energy calculations with a modest computational cost. Several approaches for this effect have been proposed in the framework of norm-conserving pseudopotentials so far and they are categorized into two approaches. One is to directly modify the behavior of the pseudopotentials, which vary sharply in the vicinity of nuclei, by filtering procedures. King-Smith {\it et al.}\cite{king-smith} were the first to introduce the careful treatment using the FF procedure for projectors of pseudopotentials in context of plane wave formalism, and several groups in quantum chemistry subsequently modified and/or simplified that filtering procedure by introducing mask functions\cite{quantchem}. These mask functions are very easy to introduce into the computational codes, because one does not need to change the codes extensively. However, the modification of pseudopotentials seriously destroys the transferability, particularly, in the procedures using mask functions\cite{quantchem}. The other approach is to reduce the grid spacing near the nuclei, and various attempts based on this approach have been made\cite{gygi,tsdg1,tsdg2,icp,multigrid}, one of which is the CPU time-saving DG technique\cite{tsdg1,tsdg2}. It is used to execute the integrals concerning the pseudopotentials on the denser grids without introducing any artificial parameter or increasing the computational costs during the self-consistency iterations, and have achieved considerable success in studying atomic configurations, electronic structures and transport properties of nanostructures\cite{ono-rsfd-appl,ono-transport,gpaw,takahashi}. Furthermore, the DG technique can be incorporated with the filtering procedures mentioned above. In the following sections, we introduce the computational scheme combining the FF proposed by King-Smith {\it et al.}\cite{king-smith} and the DG technique\cite{tsdg1,tsdg2,icp} to overcome the egg box effect in computations treating transition metals.

\subsection{Separable nonlocal form of pseudopotential}
Computations concerning wave functions and nonlocal parts of pseudopotentials are usually implemented in electronic-structure calculations using a separable form, e.g., using the procedure proposed by Kleinman and Bylander\cite{kb}. Following the notation of King-Smith {\it et al.}\cite{king-smith}, the real-space representation of a separable form may be written as
\begin{eqnarray}
v(\vecvar{r},\vecvar{r}')&=&\sum_\ell  \sum_{m=-\ell}^{+\ell} \sum_t E^t_\ell Y_{\ell m}^*(\theta_{\vecvar{r}},\phi_{\vecvar{r}})\zeta^t_\ell (r) \zeta^t_\ell (r') \\ \nonumber
&&\times Y_{\ell m}(\theta_{\vecvar{r}'},\phi_{\vecvar{r}'}),
\end{eqnarray}
where $\ell$ and $m$ are orbital and azimuthal angular momentum quantum numbers, respectively, $Y_{\ell m}$ is the spherical harmonic, $\zeta^t_\ell (r)$ is a radial projection function that vanishes outside the cutoff region of pseudopotentials ($r>r_c$), $E^t_\ell$ is an angular momentum dependent energy, and $t$ is the index of the projectors. Note that more than one projector for the same $\ell m$ are employed to improve the transferability of pseudopotentials in some cases.

Both FF and DG exploit the fact, that the pseudo wave functions are considerably smoother than the projectors of the pseudopotential. If the inner products between wave functions and the projectors are calculated naively by evaluating the integrands on each grid point and summing up these values, the required grid spacing is determined by the shape of the projectors. But a significantly coarser grid is sufficient for the precise description of the smooth wave functions, and the inner product can be rewritten as a function of wave functions that needs to be evaluated only on that coarser grid. Such a procedure can remarkably reduce the computational effort, as the operations involving the projectors but not the wave functions can be executed prior to the selfconsistency cycle. The remaining operations needed to determine the inner product are executed only for the coarse grid points.

\subsection{Fourier filtering procedure of pseudopotentials}
\label{sec:Fourier filtering procedure of pseudopotentials}
In the case of FF, the inner products between wave functions and the projectors are evaluated on a coarse grid after smoothening the projectors. This is done by transforming the projectors to Fourier space and removing the fastest-varying components as the following procedure.

$\zeta^t_\ell (r)$ is transformed to reciprocal space:
\begin{equation}
\label{eqn:qcut}
\tilde{\zeta}^t_\ell (q)= \sqrt{\frac{2}{\pi}}\int^\infty_0 r^2 \zeta^t_\ell (r) j_\ell (qr) dr,
\end{equation}
where $j_\ell$ is the spherical Bessel function of order $\ell$ and $\ell$ corresponds to the orbital angular momentum of spherical harmonics. The modified pseudopotential functions are then transformed back to real space by Fourier transform:
\begin{equation}
\chi^t_\ell(r) = \sqrt{\frac{2}{\pi}}\int^{q_{cut}}_0 q^2 \tilde{\zeta}^t_\ell(q) j_\ell(qr) dq + \Delta \chi^t_\ell(r),
\end{equation}
where $\Delta \chi^t_\ell(r)$ is an additional term such that $\chi^t_\ell(r)$ vanishes outside the sphere slightly larger than the cutoff region of pseudopotentials and $q_{cut}$ is the plane wave cutoff, which normally corresponds to $\pi/H$ with $H$ being a real-space grid spacing so as to be equal to that of the plane wave calculation that uses a Fourier transform grid with the same spacing as the RSFD calculation. We employ the original procedure proposed by King-Smith {\it et al.}\cite{king-smith} to set up $\Delta \chi^t_\ell(r)$, although many variations of the scheme have been proposed up to now\cite{quantchem}.

Thus, the expectation value of the one-particle wave function $\psi$ with respect to the projectors of pseudopotential of atom $s$ is
\begin{eqnarray}
\label{eqn:kbform}
\lefteqn{\int \int \psi^*(\vecvar{r})v^s(\vecvar{r},\vecvar{r}')\psi(\vecvar{r}') d\vecvar{r} d\vecvar{r}'} \nonumber \\
&=&\sum_\ell \sum_{m=-\ell}^{+\ell} \sum_t E^{s,t}_\ell \int \psi^*(\vecvar{r}) p^{s,t *}_{\ell m}(\vecvar{r}^s) d\vecvar{r} \int \psi(\vecvar{r}) p^{s,t}_{\ell m}({\vecvar{r}^s}) d\vecvar{r}, \nonumber \\
\end{eqnarray}
where $p^{s,t}_{\ell m}(\vecvar{r})[=Y_{\ell m}(\theta_{\vecvar{r}^s},\phi_{\vecvar{r}^s})\chi^{s,t}_\ell (r^s)]$ are the projectors, $\vecvar{r}^s=\vecvar{r}-\vecvar{R}^s$, and $\vecvar{R}^s$ is the position of the atom $s$. Furthermore, this filtering procedure is also applicable to the local parts of pseudopotentials using the separation procedure of the local parts described in Refs.~\onlinecite{icp} and \onlinecite{tsdg2}.

\subsection{Double-grid technique}
\label{sec:Time-saving double-grid technique}
In the DG technique, the inner products are evaluated on a dense grid by interpolating the pseudo wave functions $\psi(\vecvar{r})$. Since the interpolation scheme can be rewritten as a function of $\psi(\vecvar{r})$ (with $\psi(\vecvar{r})$ given on a coarse grid), the integrations are carried out on a coarse grid without degrading the numerical precision.

\begin{figure}
\begin{center}
\includegraphics{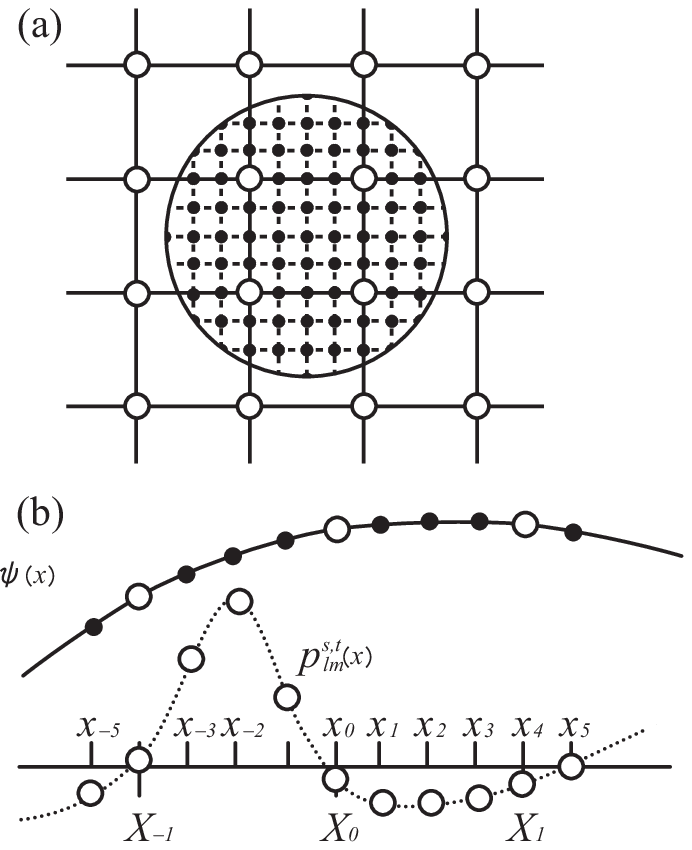}
\end{center}
\caption{(a) DG adopted in the text. The ``$\circ$'' and ``$\bullet$'' correspond to coarse- and dense-grid points, respectively. The circle shows the core region of an atom that is taken to be sufficiently large to contain the region in which the projectors are non-zero. (b) Wave function $\psi(x)$ and pseudopotential projector $p^{s,t}_{\ell m}(x)$ on coarse- and dense-grid points in the one-dimensional case. $X_J$ ($x_j$) represents a coarse- (dense-) grid point with $i=nI+\mu$ $(0 \leq \mu < n)$, and hence $X_I=x_{nI}$.}
\label{fig:1}
\end{figure}

The DG employed here consists of two types of uniform and equi-distant grid points, i.e., coarse and dense ones, depicted in Fig.~\ref{fig:1}(a) by ``$\circ$'' and ``$\bullet$'', respectively. The dense-grid region enclosed by the circle is the core region of an atom that is taken to be large enough to contain the region in which the projectors do not vanish. We postulate here that pseudo wave functions are defined and updated only on coarse-grid points, while pseudopotentials are strictly given on all dense-grid points in an analytically or numerically exact manner.

Let us consider inner products between pseudo wave functions $\psi(x)$ and the projectors $p^{s,t}_{\ell m}(x)$ [see Fig.~\ref{fig:1}(b)] in Eq.~(\ref{eqn:kbform}). In the present scheme, $p^{s,t}_{\ell m}(x)$ are the {\it filtered} projectors using the procedure in the preceding subsection while it is exactly the pseudopotential in the original DG technique. For simplicity, the illustration is limited to the one-dimensional case and an atom $s$ exits at the origin, hereafter. Pseudopotentials are made to be finite at the original, and so the resulting pseudo wave functions are rather smoothly varying functions without nodes inside the cutoff region. On the contrary, the projectors, such as the $p$-states of first-row elements and the $d$-states of transition metals, are rapidly oscillating or rapidly varying functions. In this sense, pseudo wave functions are softer than pseudopotentials. In Fig.~\ref{fig:1}(b), the values of pseudo wave functions on coarse-grid points ($\circ$) are stored in computer memory, and the values on dense-grid points ($\bullet$) are evaluated by interpolation from them. The well-known values of pseudopotentials both on coarse- and dense-grid points ($\circ$) are also shown schematically. One can see that only the values on coarse-grid points are so inadequate that the inner products cannot be precisely calculated; the errors are mainly due to the rapidly varying behavior of pseudopotentials. On the other hand, the inner products can be evaluated to great precision, if the number of dense-grid points is taken to be sufficiently large and also if the values of pseudo wave functions on dense-grid points are properly interpolated from those on coarse-grid points.

Although there are many interpolation schemes, we introduce the $\kappa$th Lagrange interpolation. The pseudo wave functions $\psi_i \equiv \psi(x_i)$ on dense-grid points $x_i$ are interpolated from $\Psi_I \equiv \psi(X_I)$ on coarse-grid points $X_I$ as
\begin{equation}
\label{eqn:inter}
\psi(x_i) = \sum_{K=-k+1}^k \Psi_{J+K} A_K(x_i)  ,
\end{equation}
where $k=[\kappa/2]+1$, $J=[i/n]$, and $A_K$ is the weight of the interpolation. In addition, $[x]$ means the maximum integer not greater than $x$. The inner product is assumed to be precisely approximated by the discrete sum over the dense-grid points, i.e.,
\begin{equation}
\label{eqn:inner}
\int^{d/2}_{-d/2} \psi(x) p^{s,t}_{\ell m}(x) dx \approx \sum_{i=-nN-n+1}^{nN+n-1} \psi(x_i) p^{s,t}_{\ell m}(x_i) h,
\end{equation}
where $d$ is the ``diameter'' of the core region, $h$ is the dense-grid spacing, and $n-1$ is the number of dense-grid points existing between adjacent coarse-grid points, i.e., $n=H/h$ with $H$ being the coarse-grid spacing, and $2N+1(2nN+1) $ is the number of coarse- (dense-) grid points in the core region. Since pseudopotentials are made to be finite, we postulate that $p^{s,t}_{\ell m}(x)$ vanishes at $|x| \ge X_{N+1} (|x| \ge x_{nN+n})$. Now, substituting Eq.~(\ref{eqn:inter}) into the right-hand side of Eq.~(\ref{eqn:inner}), we have
\begin{eqnarray}
\label{eqn:kekka}
\lefteqn{\int^{d/2}_{-d/2} \psi(x) p^{s,t}_{\ell m}(x) dx} \nonumber \\
&\approx& \sum_{i=-nN-n+1}^{nN+n-1} p^{s,t}_{\ell m}(x_i) \sum_{K=-k+1}^k \Psi_{J+K} A_K(x_i) h \nonumber \\
&=& \sum_{I=-N-1}^{N} \sum_{K=-k+1}^k \Psi_{I+K} \sum_{\mu=0}^{n-1} p^{s,t}_{\ell m}(x_{nI+\mu}) A_K(x_{nI+\mu}) h \nonumber \\
&=& \sum_{I=-N-k}^{N+k} \Psi_I w^{s,t}_{\ell m,I} H,
\end{eqnarray}
where
\begin{equation}
w^{s,t}_{\ell m,I}=\frac{1}{n} \sum_{\nu=-nk}^{nk} p^{s,t}_{\ell m}(x_{nI+\nu}) A_{-[\nu/n]}(x_{nI+\nu}).
\end{equation}
As shown in Eq.~(\ref{eqn:kekka}), the right-hand side of the inner product Eq.~(\ref{eqn:inner}) has been replaced with the summation over coarse-grid points inside the core region, which produces only a modest overhead in the computational cost.

In the PAW formalism, the local effective potential (sum of Coulomb and exchange-correlation potentials) is described on a grid that is two or three times denser than that for pseudo wave functions. This is a particular difference of the PAW method from the norm-conserving pseudopotentials. To solve the eigenvalue problem for the Kohn-Sham Hamiltonian, the local effective potential has to be described on the coarse grid points because pseudo wave functions $\Psi_I$ are defined only on the coarse grid. This transformation from the dense grid to the coarse grid is achieved by the Fourier transform in the case of the plane wave method. On the other hand, in the RSFD scheme, the egg box effect is a serious problem if one simply picks up the values of the potential on the dense grid and uses them in the eigenvalue problem. Assuming that the dense grid for the local effective potential of the PAW method corresponds to that in the DG technique, we can apply the DG technique to the transformation of the local effective potential as introduced in Refs.~\onlinecite{icp} and \onlinecite{tsdg2}. In the energy functional, the inner product of the local effective potential $v^{eff}(x)$ and pseudo charge density $\rho(x)$, which is defined as $\tilde{n}(r)$ in Ref.~\onlinecite{paw}, is given by

\begin{equation}
\label{eqn:innercc}
\int_\Omega \rho(x) v^{eff}(x) dx \approx \sum_{i=-nN_{all}-n+1}^{nN_{all}+n-1} \rho(x_i) v^{eff}(x_i) h,
\end{equation}
where $\Omega$ and $2N_{all}+1$ are the volume of the computational region and the number of coarse grid points in the computational region, respectively. The pseudo charge density $\rho_i \equiv n(x_i)$ on dense-grid points $x_i$ is interpolated from $P_I \equiv \rho(X_I)$ on coarse-grid points $X_I$ as
\begin{equation}
\label{eqn:intercc}
\rho(x_i) = \sum_{K=-k+1}^k P_{J+K} A_K(x_i).
\end{equation}
By substituting Eq.~(\ref{eqn:intercc}) into the right-hand side of Eq.~(\ref{eqn:innercc}) and taking into account that the smooth pseudo charge density vanishes outside the computational region in the case of an isolated boundary condition, we have
\begin{eqnarray}
\label{eqn:kekkacc}
\lefteqn{\int_\Omega \rho(x) v^{eff}(x) dx} \nonumber \\
&\approx& \sum_{i=-nN_{all}-n+1}^{nN_{all}+n-1} v^{eff}(x_i) \sum_{K=-k+1}^k P_{J+K} A_K(x_i) h \nonumber \\
&=& \sum_{I=-N_{all}-1}^{N_{all}} \sum_{K=-k+1}^k P_{I+K} \sum_{\mu=0}^{n-1} v^{eff}(x_{nI+\mu}) A_K(x_{nI+\mu}) h \nonumber \\
&=& \sum_{I=-N_{all}}^{N_{all}} P_I w^{eff}_I H,
\end{eqnarray}
where
\begin{equation}
w^{eff}_I=\frac{1}{n} \sum_{\nu=-nk}^{nk} v^{eff}(x_{nI+\nu}) A_{-[\nu/n]}(x_{nI+\nu}).
\end{equation}
Calculating the derivative with respect to the pseudo wave functions $\Psi_I$, we find that $w^{eff}_I$ is the contribution of the local effective potential in the eigenvalue problem for the Kohn-Sham Hamiltonian. The extension to a periodic boundary conditions is straightforward.

\section{Performance results}
\label{sec:Performance}
We now examine the efficiency of the present combination of the FF and the DG technique. Hereafter, we choose the seventeen-point finite-difference formula, i.e., $N_f=8$ in Eq.~(1) of Ref.~\onlinecite{chelikowsky2}, for the differentiation of the wave function. The dense-grid spacing is set as $h_\mu =H_\mu/3$, where $H_\mu$ ($\mu=x, y,$ and $z$) is the coarse-grid spacing in the $\mu$ direction. The seventeenth-order Lagrangian interpolation is used for the interpolation of the DG technique. In order to demonstrate the precision of the RSFD calculations, our results are compared to those obtained by PWPAW\cite{kresse-joubert} and FLAPW\cite{flapw,fleur} calculations. In all calculations throughout the paper we used the Vosko, Wilk and Nussair\cite{VWN-lda} approximation to treat the unknown exchange correlation functional within the framework of the density-functional theory\cite{dft}. The parameters for the pseudopotential generation and the PAW data sets for the RSFD calculation are summarized in Table~\ref{tbl:1}, which are taken from another first-principles code based on the plane wave formalism\cite{estcompp}. All cut-off parameters for the RSFD and FLAPW calculations are given with the respective data presented and the data obtained by the PWPAW method are taken from Ref.~\onlinecite{kresse-joubert}.

\begin{table}
\caption{Parameters of PAW data sets used in the present work. $r_c$, $q_{cut}$, and $\varepsilon^t_l$ ($l$=$s$, $p$, and $d$, and $t$=1 and 2) are the cutoff radius, the filtering parameter in Eq.~(\ref{eqn:qcut}), and the eigenvalue of the partial waves, respectively. $R_0$ and $\gamma$ are the filtering parameters defined in Ref.~\onlinecite{king-smith}. $\varepsilon^t_l$ in the first and second lines is the reference energy for the first ($t$=1) and second ($t$=2) projectors, respectively, and $\varepsilon^1_l$ corresponds to the eigenvalue of the Kohn-Sham Hamiltonian.}
\begin{ruledtabular}
\begin{tabular}{lccccccc}
& $r_c$ (bohr) & $R_0$ & $q^2_{cut}$ (Ry)& $\gamma^2$ (Ry) & $\varepsilon^t_s$ (Ry) & $\varepsilon^t_p$ (Ry) & $\varepsilon^t_d$ (Ry) \\
\hline
Cu &$2.20$& 1.2$r_c$ &$25$& 100 &$-0.358$&$-0.058$&$-0.392$ \\
   &      &          &    &     &$ 1.642$&        &$ 0.208$ \\
Fe &$2.10$& 1.2$r_c$ &$25$& 100 &$-0.402$&$-0.106$&$-0.570$ \\
   &      &          &    &     &        &        &$ 0.030$ \\
Ni &$2.15$& 1.2$r_c$ &$25$& 100 &$-0.430$&$-0.098$&$-0.672$ \\
   &      &          &    &     &$      $&$      $&$ 0.228$ \\
Co &$2.10$& 1.2$r_c$ &$25$& 100 &$-0.416$&$-0.102$&$-0.622$ \\
   &      &          &    &     &        &        &$-0.022$ \\
Ti &$2.25$& 1.2$r_c$ &$25$& 100 &$-0.338$&$-0.113$&$-0.328$ \\
   &      &          &    &     &        &        &$ 0.072$ \\
\end{tabular}
\end{ruledtabular}
\label{tbl:1}
\end{table}

\subsection{Total energy convergence}
\begin{figure}
\begin{center}
\includegraphics{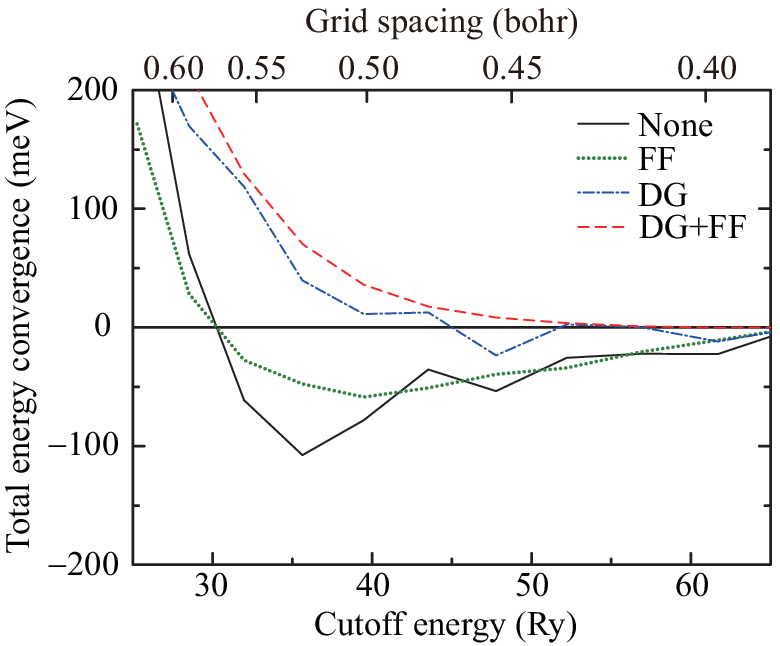}
\end{center}
\caption{(color online) Convergence of total energy for Cu atom as a function of the coarse-grid cutoff energy $\pi^2/H^2$ (Ry). Zero total energy is set at that computed with cutoff energy of 67 Ry.}
\label{fig:2}
\end{figure}

The total energy convergence with respect to grid spacing is an important test for the RSFD formalism utilizing the present combination of the FF and the DG technique because computations with large grid spacing are one of the advantages of the PAW pseudopotentials over norm-conserving ones and grid spacing is closely relevant to computational cost. An isolated Cu atom is selected as an example in this test. The Cu atom is placed in the center of the neighboring grid points for the $x$, $y$ and $z$ directions. The total energy convergence as a function of the cutoff energy is depicted in Fig.~\ref{fig:2}. In accordance with Ref. \onlinecite{gygi}, we defined a cutoff energy, $\pi^2/H_\mu^2$ (Ry), that is equivalent to that of the plane wave formalism which uses a fast-Fourier-transform grid with the same spacing as the present calculation. When neither the FF nor the DG is used, the total energy does not converge even when the cutoff energy increases to 60 Ry. Although we can also obtain good convergence when either the FF or the DG is adopted, one can see that the use of the combination of the FF and the DG yields the best convergence among them; convergence to 1 meV/atom is achieved at about 55 Ry.

\subsection{Egg box effect}
\begin{figure}
\begin{center}
\includegraphics{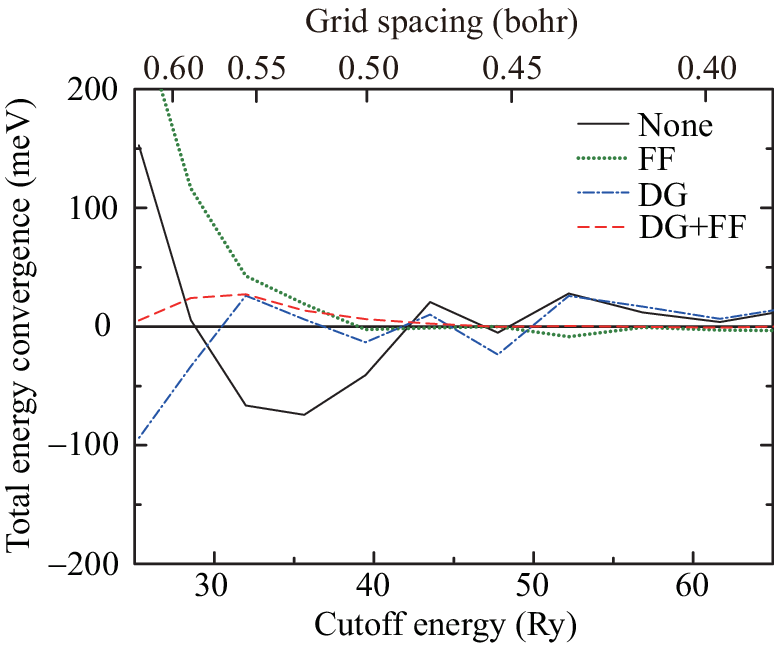}
\end{center}
\caption{(color online) Total energy variation $E-E'$ of Cu atom owing to egg box effect as a function of the coarse-grid cutoff energy $\pi^2/H^2$ (Ry). Here, $E$ is the total energy when the atom is located at the center between adjacent coarse-grid points and $E'$ is the energy when the atom is shifted by 0.5$H_x$ along the $x$ direction.}
\label{fig:3}
\end{figure}

To illustrate the efficiency of the present scheme for the egg box effect, a test calculation is performed on the total energy variation with respect to grid points. In the calculation using a discrete grid, the loss of translational invariance manifests spurious variation of the total energies and forces, which prevents us from implementing practical calculations. Fig.~\ref{fig:3} shows the difference in the total energy between the cases in which the atom is placed in the center of the neighboring grid and the atom is shifted by 0.5 $H_x$ along the $x$ axis as a function of the cutoff energy. The loss of translational invariance in the present combination of the FF and the DG technique is the smallest and the difference at about 45 Ry is as small as 1 meV/atom. In our previous study\cite{tsdg2}, the interatomic distance of Cu dimer were examined using the norm-conserving pseudopotentials. The grid spacing was 0.265 bohr and the deviation of the total energy due to the egg box effect was $\sim$4 meV/atom. From the present results concerning the total energy convergence and the prescription for the egg box effect, a grid spacing of 0.42 bohr is sufficient for the precise treatments of the system including Cu in the present scheme, which means that we have succeeded to reduce the number of grid points 75\% and improve the precision compared to our previous scheme.

\subsection{Consistency between total energy and force}
\label{sec:consist}
\begin{figure}
\begin{center}
\includegraphics{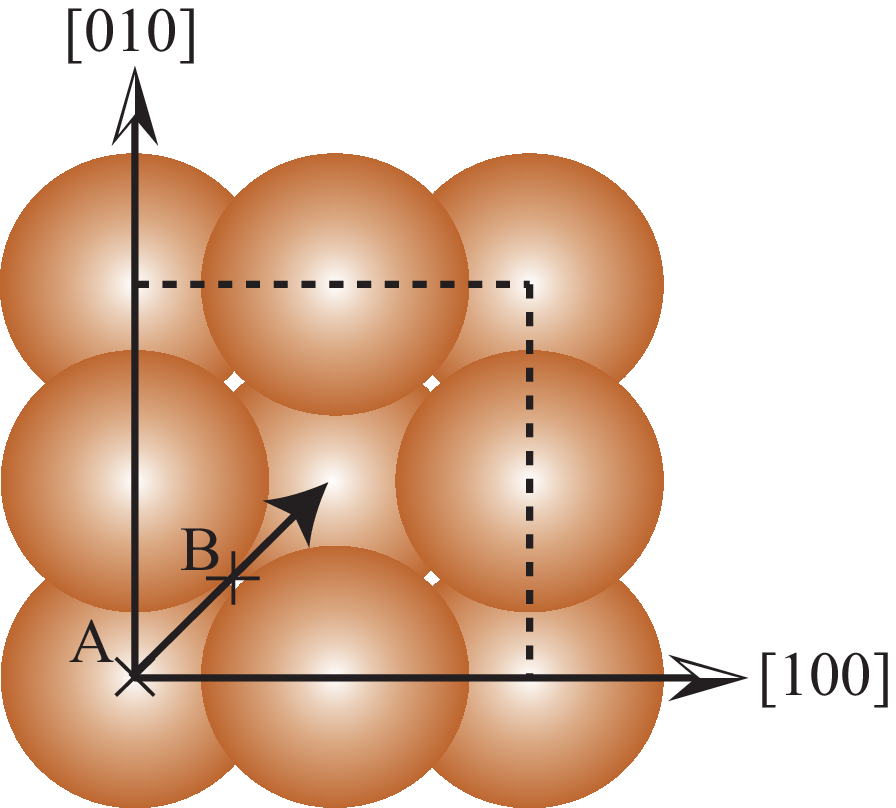}
\end{center}
\caption{(color online) Top view of Cu(001) surface-layer-atoms, second-layer atoms and adatom (crosses) for the check of consistency between total energy and force discussed in the text. The test system is a cell of 5 Cu atoms consisting of one adatom and two rigid Cu(001) planes. This model is also employed to compute the energy barrier for the surface migration in Sec.~\ref{sec:migration}.}
\label{fig:4}
\end{figure}

\begin{figure}
\begin{center}
\includegraphics{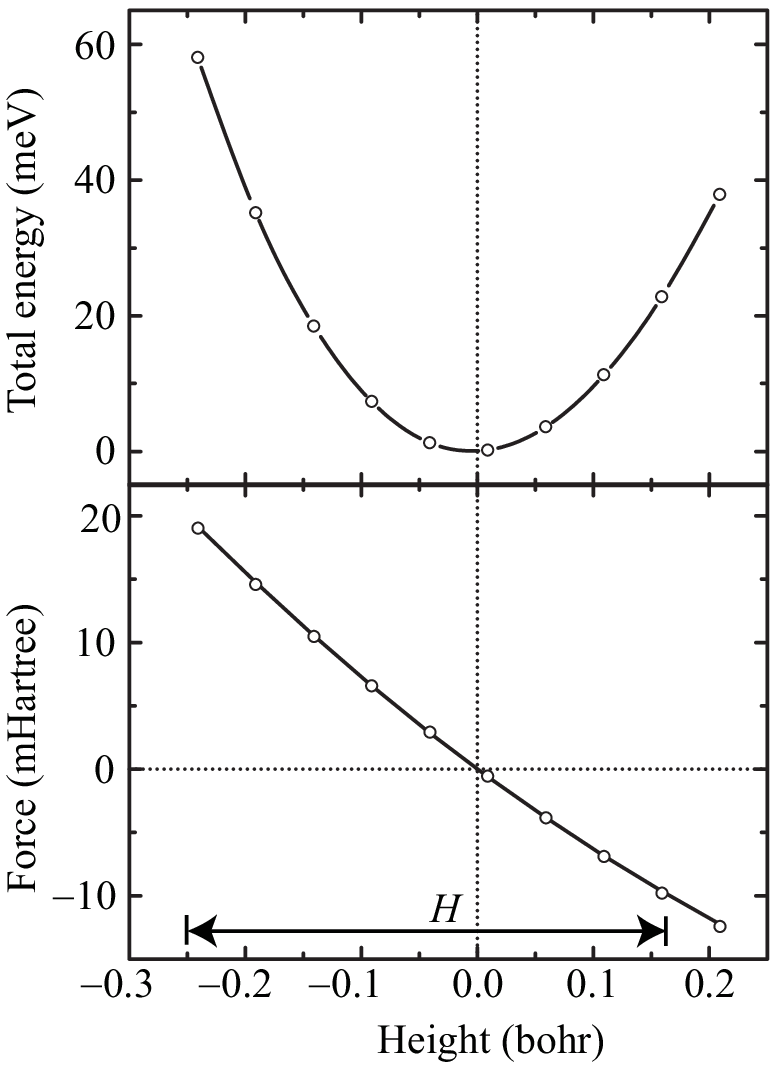}
\end{center}
\caption{Energy and force for an adatom on Cu(001) surface at different heights. Top: The circles show the calculated energies and the curve shows a spline fit. Bottom: The circles show the calculated forces and the curve is minus the derivative of the energies in the top panel using the seven-points finite-difference formula. The zero of the height is set at the most stable position. The edges of the arrows in the lower panel correspond to the coarse grid planes parallel to the Cu(001) surface.}
\label{fig:5}
\end{figure}

Consistency between total energy and force is of importance for molecular-dynamics simulations. We then compute the force acting on atoms to check the consistency. Figure~\ref{fig:4} shows the computational model: The supercell, which contains 5 Cu atoms consisting of one adatom and two rigid Cu(001) planes, is $L_x$=$L_y$=$a_0$ and $L_z$=4$a_0$ under periodic boundary condition, where $L_x$, $L_y$, and $L_z$ are the lengths of the supercell in the $x$, $y$, and $z$ directions, respectively, and $a_0$ is the experimental lattice constant of the Cu bulk (6.68 bohr). A grid spacing of 0.42 bohr, which corresponds to the plane wave cutoff energy of 57 Ry, is used and 8$\times$8$\times$1 $\vecvar{k}$-point mesh is employed for the integration in the Brillouin zone. The position of the adatom is displaced by 0.05 bohr along the $z$ axis. The total energy and force as a function of the height of the adatom from the surface are plotted in Fig. ~\ref{fig:5}. The numerical derivation of the total energy is computed using the seven-points finite-difference formula; the error due to the numerical derivation is $\sim$ 10$^{-8}$ mHartree in this case. The maximum difference between the computed force and the numerical derivation is 0.17 mH/bohr. Note that the consistency between total energy and force is excellent and the precision of the force is enough to implement first-principles molecular-dynamics simulations. 

\subsection{Migration energy}
\label{sec:migration}
To demonstrate the precision of force in the practical calculations, the energy barrier for a Cu adatom to hop along the [110] direction on the Cu(001) surface is examined using the same surface with that of Sec.~\ref{sec:consist}. We evaluate the ground-state geometry for the adatom located at a hollow site A, and then displace the adatom along the [110] direction from the fcc hollow site A to the nearest hollow site via the bridge site B (see Fig.~\ref{fig:4}). For comparison, we compute the energy barrier using FLAPW\cite{flapw,fleur} and plane wave\cite{state} calculations. The plane wave method employs the ultrasoft pseudopotentials proposed by Vanderbilt\cite{vanderbilt}. Same $\vecvar{k}$-point set and supercell are adopted in all three calculations. The optimized height of the adatom from the surface layer and computed migration energy barrier are shown in Table~\ref{tbl:migration}. The numerical errors are found to be 0.2\% and 0.3\% for the height of the adatom and migration energy, respectively, and those are negligible in practical simulations.

\begin{table}
\caption{Migration energy and height from surface layer of adatom on Cu(001). $\Delta Z_A$ and $\Delta Z_B$ are the height of the adatom at the positions A and B in Fig.~\ref{fig:4}, respectively. Plane wave cutoff energy of 25.0 Ry and muffin-tin radius of 2.1 bohr are employed in the FLAPW calculations\cite{fleur}. Plane wave cutoff energy of 36.0 Ry and cutoff radius of 2.2 bohr for Vanderbilt's ultrasoft pseudopotentials are employed in the plane wave calculations\cite{state}.}
\begin{ruledtabular}
\begin{tabular}{lccc}
& Migration energy (meV) & $\Delta Z_A$ (bohr) & $\Delta Z_B$ (bohr) \\
\hline
RSFD  & 661 & 2.97 & 3.61 \\
FLAPW & 673 & 2.97 & 3.61 \\
Plane wave method & 672 & 2.96 & 3.60 \\
\end{tabular}
\end{ruledtabular}
\label{tbl:migration}
\end{table}

\subsection{Bulk properties}
Our final series of tests is the calculation of the bulk properties of transition metals. These systems are usually treated with the plane-wave basis set. In order to ensure the precision of the RSFD formalism with the present combination of the FF and the DG, the computed bulk properties are compared with other theoretical results. The cuboid supercells are employed and the {\it k}-space integrations are performed with 15 $\times$ 15 $\times$ 15 \vecvar{k}, 12 $\times$ 12 $\times$ 12 \vecvar{k}, and 14 $\times$ 8 $\times$ 8 \vecvar{k} points, yielding 240, 240, and 185 $k$ points in the irreducible wedge of the Brillouin zone for the body-centered-cubic (bcc), fcc, and hcp structures, respectively. Table~\ref{tbl:2} clearly shows that our results are in excellent agreement with those obtained using the PWPAW and FLAPW codes. The parameters in the PAW pseudopotentials affect the results slightly, therefore the difference in the results between the three numerical methods are mainly attributed by the PAW pseudopotentials.

\begin{table}
\caption{Equilibrium lattice constant ($a$ and $c$), bulk modulus ($B$), and magnetic moment ($M_0$). The grid spacings (the cutoff energies) of the RSFD code are 0.43, 0.41, 0.42, 0.42, and 0.40 bohr (52, 60, 57, 57, and 62 Ry) for Fe, Ni, Co, Cu, and Ti, respectively. In the FLAPW calculations\cite{fleur}, we use plane wave cutoff energies of 25.0 Ry (Fe, Ni, and Cu) and 14.4 Ry (Co and Ti), and muffin-tin radii of 2.1$-$2.2 bohr (Fe, Ni, Co, and Cu) and 2.6 bohr (Ti).}
\begin{ruledtabular}
\begin{tabular}{lccccc}
&Ref.&$a$ (bohr)& $c/a$ & B(Mbar) & $M_0$ ($\mu_B$) \\
\hline
bcc Fe &&&&& \\
\hspace{5mm}RSFD  &&5.21&---&2.53&1.97 \\
\hspace{5mm}PWPAW &\onlinecite{kresse-joubert}&5.20&---&2.47&2.00 \\
\hspace{5mm}FLAPW &\onlinecite{fleur}&5.20&---&2.54&2.00 \\
fcc Ni &&&&& \\
\hspace{5mm}RSFD  &&6.48&---&2.63&0.58 \\
\hspace{5mm}PWPAW &\onlinecite{kresse-joubert}&6.48&---&2.51&0.58 \\
\hspace{5mm}FLAPW &\onlinecite{fleur}&6.47&---&2.63&0.59 \\
hcp Co &&&&& \\
\hspace{5mm}RSFD  &&4.59&1.61&---&1.51 \\
\hspace{5mm}PWPAW &\onlinecite{kresse-joubert}&4.59&1.62&---&1.51 \\
\hspace{5mm}FLAPW &\onlinecite{fleur}&4.59&1.61&---&1.52 \\
fcc Cu &&&&& \\
\hspace{5mm}RSFD  &&6.65&---&1.85&0.00 \\
\hspace{5mm}FLAPW &\onlinecite{fleur}&6.65&---&1.90&0.00 \\
hcp Ti &&&&& \\
\hspace{5mm}RSFD  &&5.42&1.58&---&0.00 \\
\hspace{5mm}FLAPW &\onlinecite{fleur}&5.42&1.58&---&0.00 \\
\end{tabular}
\end{ruledtabular}
\label{tbl:2}
\end{table}

We next compute the structural energy differences between different bulk phases of transition metals. This calculation requires an extremely high precision because the differences are quite small. The cutoff energies and the number of sampling $\vecvar{k}$-points are the same as those listed in Table~\ref{tbl:2}. As can be seen from Table~\ref{tbl:3}, the agreement between the current RSFD code implemented in the present scheme, the PWPAW code and the FLAPW code is excellent. With the exception of Fe all ground state structures are well reproduced with deviations in the order of 10\% or better. This is really an achievement considering the difference of the methods. Even the fcc structure of Cu is well reproduced which is only about 8 meV lower in energy than the hcp structure. Fe deserves a special mentioning. Our calculations do not reproduce the bcc ground-state structure, a well-known failure of the local-spin-density approximation\cite{VWN-lda} and cannot be not attributed to the computational methods. In fact the results obtained by the different methods are internally consistent. These results imply that the RSFD formalisms with the combination of the FF and the DG technique are readily applicable to simulations treating transition metals with a high degree of precision.

\begin{table}
\caption{Structural energy differences between different bulk phases at the equilibrium lattice constants. The most stable phase is chosen as zero energy. Unit is meV/atom. For hcp Fe, $c/a$ is set at ideal value to keep the consistency with Ref. \onlinecite{kresse-joubert}. Most computational parameters as in Table~\ref{tbl:2}, but in the FLAPW calculations we increased the cutoff energies to 27.0 Ry and 25.0 Ry for Co and Ti, respectively.}
\begin{ruledtabular}
\begin{tabular}{lccccc}
&Ref.&hcp& fcc & NM bcc & FM bcc \\
\hline
Fe &&&&& \\
\hspace{5mm}RSFD  &&$-$142&$-$72&+288&0 \\
\hspace{5mm}PWPAW &\onlinecite{kresse-joubert}&$-$139&$-$68&+282&0 \\
\hspace{5mm}FLAPW &\onlinecite{fleur}&$-$142&$-$68&+287&0 \\
Co &&&&& \\
\hspace{5mm}RSFD  &&0&+20&---&--- \\
\hspace{5mm}FLAPW &\onlinecite{fleur}&0&+23&---&--- \\
Cu &&&&& \\
\hspace{5mm}RSFD  &&+9&0&---&--- \\
\hspace{5mm}FLAPW &\onlinecite{fleur}&+8&0&---&--- \\
Ti &&&&& \\
\hspace{5mm}RSFD  &&0&+53&---&--- \\
\hspace{5mm}FLAPW &\onlinecite{fleur}&0&+54&---&--- \\
\end{tabular}
\end{ruledtabular}
\label{tbl:3}
\end{table}

\section{Summary and conclusion}
We developed a first-principles electronic structure method that solves the projector augmented wave (PAW) formalism by a real-space finite-difference (RSFD) approach which exhibits full-potential all-electron precision with a moderate grid spacing even for transition metals. The scheme developed combines the advantage of the time-saving double-grid technique with the capability of the Fourier-filtering procedure and was benchmarked agains the full-potential linearized augmented plane wave (FLAPW) method. The FLAPW method is widely considered the most accurate first-principles method for solids, which essentially provides the Kohn-Sham answer to the chosen functional approximating the unknown exchange correlation energy. We have shown that the present RSFD method is a band-structure method that gives ground state properties, such as lattice parameters, bulk moduli, and magnetic moments with the same accuracy as the FLAPW method, If identical exchange correlation functionals are used in the calculations, the differences between the RSFD and the FLAPW codes results are as small as $\sim$ 0.01 bohr, $\sim$ 0.05 Mbar, and $\sim$ 0.03 $\mu_B$ for lattice parameters, bulk modulii, and magnetic moments, respectively. Although the cutoff energy required in the RSFD calculations is almost twice as large as that employed in the plane wave formalisms, this disadvantage can be compensated by the advantages of the RSFD formalism, e.g., excellent performance on massively parallel computers, small computational cost for the operations concerning the projectors of pseudopotentials, and suitable algorithms for linear scaling computation. The development of a simulation code combining the RSFD formalism and the PAW method dedicated to performing large-scale first-principles calculations on massively parallel computers is in progress.

\section*{Acknowledgments}
The authors would like to thank Kikuji Hirose and Yoshitada Morikawa of Osaka University, and Ionut Tranca and Daniel Wortmann of Forschungszentrum J\"ulich for fruitful discussion. This research was partially supported by Strategic Japanese-German Cooperative Program from Japan Science and Technology Agency and Deutsche Forschungsgemeinschaft, by a Grant-in-Aid for Young Scientists (B) (Grant No. 20710078), and also by a Grant-in-Aid for the Global COE ``Center of Excellence for Atomically Controlled Fabrication Technology'' from the Ministry of Education, Culture, Sports, Science and Technology, Japan. T.O. thanks the Alexander von Humboldt Foundation and N.A. and P.B. thank the Japan Society for the Promotion of Science for the financial support. The numerical calculation was carried out using the computer facilities of the Institute for Solid State Physics at the University of Tokyo, the Research Center for Computational Science at the National Institute of Natural Science, Center for Computational Sciences at University of Tsukuba, the Information Synergy Center at Tohoku University, and Supercomputing Centre at Forschungszentrum J\"ulich.

\end{document}